\documentclass[fleqn,a4paper,12pt,twoside]{article}
\usepackage{espcrc1}

\usepackage{graphicx}
\usepackage[figuresright]{rotating}

\newcommand{\rb}[1]{\mbox{\rm \scriptsize #1}}
\newcommand{\rbb}[1]{\mbox{\rm #1}}

\title{Results on Correlations and Fluctuations from NA49}

\author{C. Blume for the NA49 collaboration \\
\vspace{0.3cm}
\begin{footnotesize}
S.V.~Afanasiev$^{9}$,T.~Anticic$^{21}$, B.~Baatar$^{9}$,D.~Barna$^{5}$,
J.~Bartke$^{7}$, R.A.~Barton$^{3}$, M.~Behler$^{15}$,
L.~Betev$^{10}$, H.~Bia{\l}\-kowska$^{19}$, A.~Billmeier$^{10}$,
C.~Blume$^{8}$, C.O.~Blyth$^{3}$, B.~Boimska$^{19}$, M.~Botje$^{1}$,
J.~Bracinik$^{4}$, R.~Bramm$^{10}$, R.~Brun$^{11}$,
P.~Bun\v{c}i\'{c}$^{10,11}$, V.~Cerny$^{4}$, O.~Chvala$^{17}$,
J.G.~Cramer$^{18}$, P.~Csat\'{o}$^{5}$, P.~Dinkelaker$^{10}$,
V.~Eckardt$^{16}$, P.~Filip$^{16}$,
H.G.~Fischer$^{11}$, Z.~Fodor$^{5}$, P.~Foka$^{8}$, P.~Freund$^{16}$,
V.~Friese$^{8,15}$, J.~G\'{a}l$^{5}$,
M.~Ga\'zdzicki$^{10}$, G.~Georgopoulos$^{2}$, E.~G{\l}adysz$^{7}$,
S.~Hegyi$^{5}$, C.~H\"{o}hne$^{15}$, G.~Igo$^{14}$,
P.G.~Jones$^{3}$, K.~Kadija$^{11,21}$, A.~Karev$^{16}$,
V.I.~Kolesnikov$^{9}$, T.~Kollegger$^{10}$, M.~Kowalski$^{7}$,
I.~Kraus$^{8}$, M.~Kreps$^{4}$, M.~van~Leeuwen$^{1}$,
R.~Lednick\'y$^{16}$,
P.~L\'{e}vai$^{5}$, A.I.~Malakhov$^{9}$, S.~Margetis$^{13}$,
C.~Markert$^{8}$, B.W.~Mayes$^{12}$, G.L.~Melkumov$^{9}$,
C.~Meurer$^{10}$,
A.~Mischke$^{8}$, M.~Mitrovski$^{10}$,
J.~Moln\'{a}r$^{5}$, J.M.~Nelson$^{3}$,
G.~P\'{a}lla$^{5}$, A.D.~Panagiotou$^{2}$,
K.~Perl$^{20}$, A.~Petridis$^{2}$, M.~Pikna$^{4}$, L.~Pinsky$^{12}$,
F.~P\"{u}hlhofer$^{15}$,
J.G.~Reid$^{18}$, R.~Renfordt$^{10}$, W.~Retyk$^{20}$,
C.~Roland$^{6}$, G.~Roland$^{6}$, A.~Rybicki$^{7}$, T.~Sammer$^{16}$,
A.~Sandoval$^{8}$, H.~Sann$^{8}$, N.~Schmitz$^{16}$, P.~Seyboth$^{16}$,
F.~Sikl\'{e}r$^{5}$, B.~Sitar$^{4}$, E.~Skrzypczak$^{20}$, J.~Smolik$^{16}$,
G.T.A.~Squier$^{3}$, R.~Stock$^{10}$, H.~Str\"{o}bele$^{10}$, T.~Susa$^{21}$,
I.~Szentp\'{e}tery$^{5}$, J.~Sziklai$^{5}$,
T.A.~Trainor$^{18}$, D.~Varga$^{5}$, M.~Vassiliou$^{2}$,
G.I.~Veres$^{5}$, G.~Vesztergombi$^{5}$,
D.~Vrani\'{c}$^{8}$, S.~Wenig$^{11}$, A.~Wetzler$^{10}$, C.~Whitten$^{14}$,
I.K.~Yoo$^{8,15}$, J.~Zaranek$^{10}$, J.~Zim\'{a}nyi$^{5}$ \\
\end{footnotesize}
\vspace{0.2cm}
\begin{scriptsize}
\noindent
$^{1}$NIKHEF, Amsterdam, Netherlands. \\
$^{2}$Department of Physics, University of Athens, Athens, Greece.\\
$^{3}$Birmingham University, Birmingham, England.\\
$^{4}$Comenius University, Bratislava, Slovakia.\\
$^{5}$KFKI Research Institute for Particle and Nuclear Physics, Budapest, Hungary.\\
$^{6}$MIT, Cambridge, USA.\\
$^{7}$Institute of Nuclear Physics, Cracow, Poland.\\
$^{8}$Gesellschaft f\"{u}r Schwerionenforschung (GSI), Darmstadt, Germany.\\
$^{9}$Joint Institute for Nuclear Research, Dubna, Russia.\\
$^{10}$Fachbereich Physik der Universit\"{a}t, Frankfurt, Germany.\\
$^{11}$CERN, Geneva, Switzerland.\\
$^{12}$University of Houston, Houston, TX, USA.\\
$^{13}$Kent State University, Kent, OH, USA.\\
$^{14}$University of California at Los Angeles, Los Angeles, USA.\\
$^{15}$Fachbereich Physik der Universit\"{a}t, Marburg, Germany.\\
$^{16}$Max-Planck-Institut f\"{u}r Physik, Munich, Germany.\\
$^{17}$Institute of Particle and Nuclear Physics, Charles University, Prague, Czech Republic.\\
$^{18}$Nuclear Physics Laboratory, University of Washington, Seattle, WA, USA.\\
$^{19}$Institute for Nuclear Studies, Warsaw, Poland.\\
$^{20}$Institute for Experimental Physics, University of Warsaw, Warsaw, Poland.\\
$^{21}$Rudjer Boskovic Institute, Zagreb, Croatia.\\
\end{scriptsize}
}
       
\begin{document}

\maketitle

\begin{abstract}

The large acceptance and high momentum resolution as well as the significant particle
identification capabilities of the NA49 experiment\cite{NA49NM} at the CERN SPS
allow for a broad study of fluctuations and correlations in hadronic interactions.

In the first part recent results on event-by-event charge and 
$\langle p_{\rb{t}} \rangle$ fluctuations are presented. 
Charge fluctuations in central Pb+Pb reactions are investigated 
at three different beam energies (40, 80, and 158 $A$GeV), while 
for the $\langle p_{\rb{t}} \rangle$ fluctuations the focus is put on
the system size dependence at 158 $A$GeV.

In the second part recent results on Bose Einstein correlations of 
$\rbb{h}^{-}\rbb{h}^{-}$ pairs in 
minimum bias Pb+Pb reactions at 40 and 158 $A$GeV, as well as of K$^{+}$K$^{+}$ and
K$^{-}$K$^{-}$ pairs in central Pb+Pb collisions at 158 $A$GeV are shown.
Additionally, other types of two particle correlations, namely $\pi$p, $\Lambda$p, 
and $\Lambda \Lambda$ correlations, have been measured by the NA49 experiment.
Finally, results on the energy and system size dependence of deuteron
coalescence are discussed.

\end{abstract}

\section{Fluctuations}

\subsection{Charge fluctuations}

\begin{figure}[h]
\begin{minipage}[b]{77mm}
\begin{center}
\includegraphics[height=73mm]{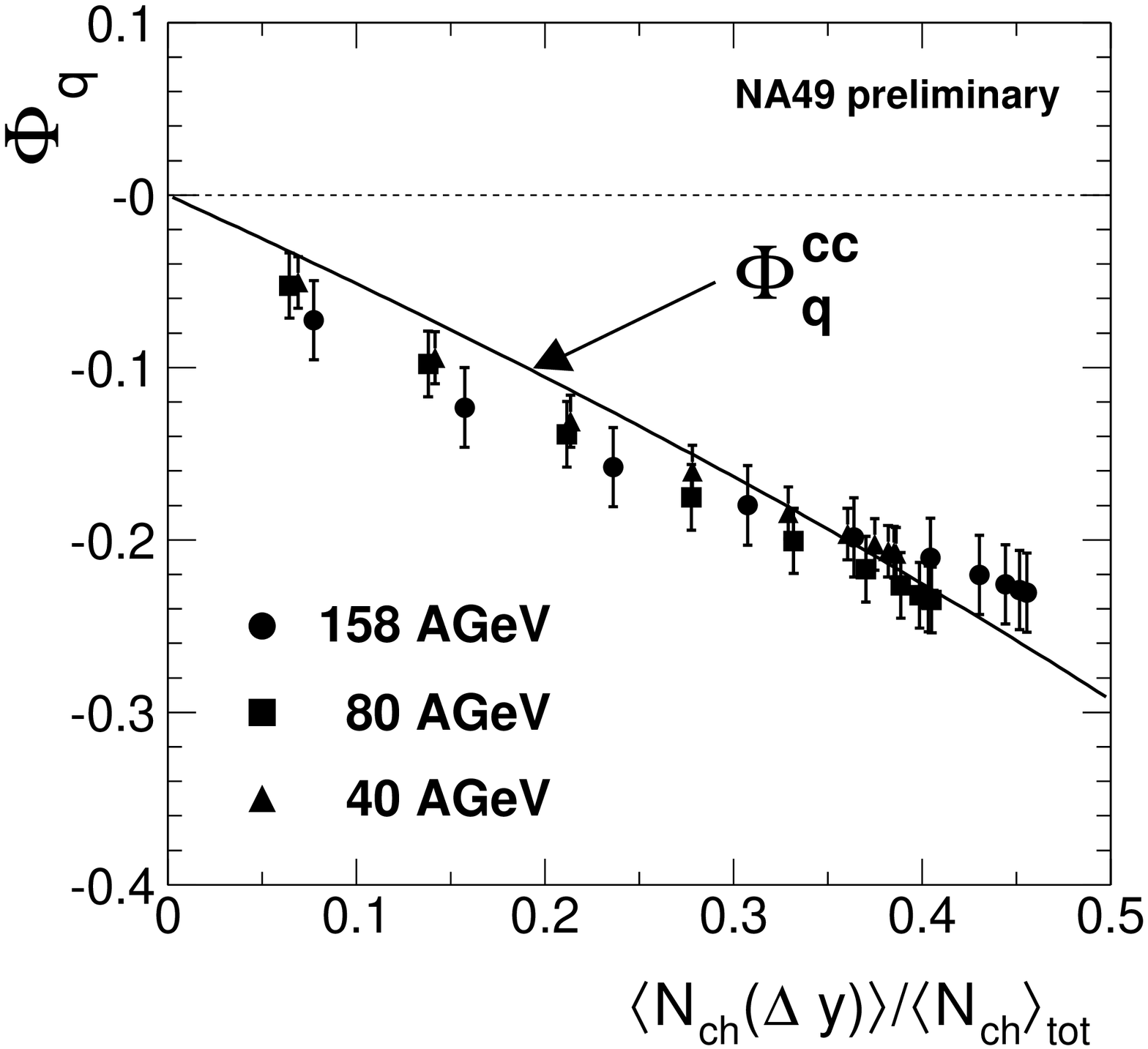}
\end{center}
\vspace{-1.2cm}
\caption{$\Phi_{\rb{q}}$ as a function of the number of charged particles in the
acceptance window $\Delta y$ for central Pb+Pb collisions.}
\label{fig:phiq}
\end{minipage}
\hspace{\fill}
\begin{minipage}[b]{77mm}
\begin{center}
\includegraphics[height=73mm]{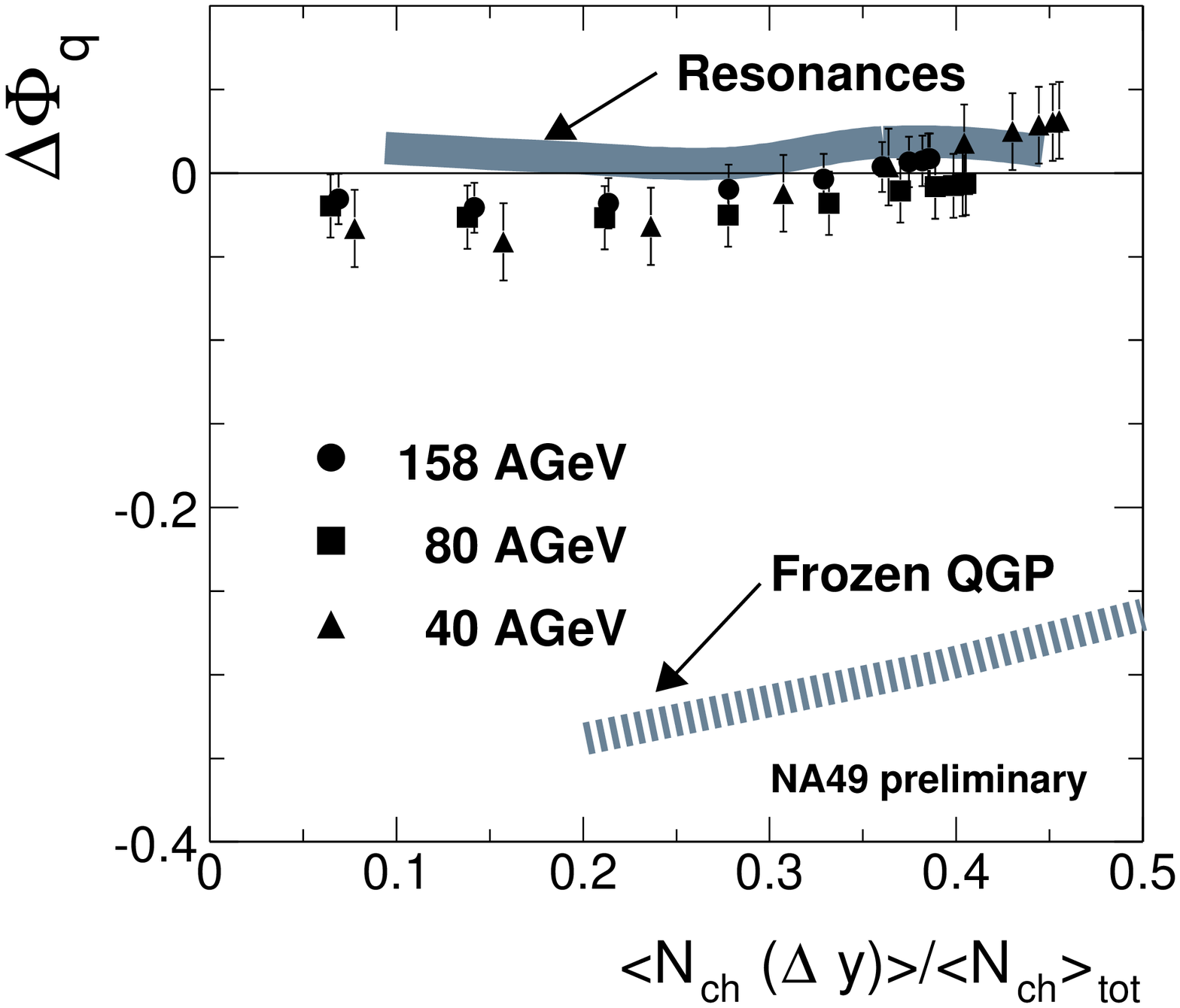}
\end{center}
\vspace{-1.2cm}
\caption{$\Delta \Phi_{\rb{q}}$ as a function of the number of charged particles in the
acceptance window $\Delta y$ for central Pb+Pb collisions.}
\label{fig:dphiq}
\end{minipage}
\end{figure}

It has been suggested that charge fluctuations might be sensitive to the
presence of a quark qluon plasma phase \cite{Asak00,Jeon00}.
A suitable observable, however, must take into account the effect of
impact parameter fluctuations, acceptance effects, the total net charge
of the reaction system, and charge conservation. For this purpose a
generalized $\Phi_{\rb{x}}$ measure \cite{Gazd92} can be employed:
\begin{eqnarray}
  \Phi_{\rb{x}} = \sqrt{\langle Z^{2} \rangle / \langle N \rangle}
                - \sqrt{\bar{z^{2}}} &
  \rbb{with:}\;
  Z = \sum^{N}_{i = 1} z_{i}  &
  \rbb{and} \;
  z_{i} =  x_{i} - \bar{x} 
\label{eq:phix}
\end{eqnarray}
Here $N$ is the event multiplicity, while $x$  is the quantity to be
studied.
The overline denotes the average over a single particle inclusive
distribution,
whereas $\langle \cdots \rangle$ means the average over all events.
In the case of charge fluctuations, $x$ is chosen as the electric charge
$q$ of a particle \cite{Zara01}.
$\Phi_{\rb{q}}$ can vary between two extreme cases: For independent
particle emission $\Phi_{\rb{q}}$ is equal to 0, while local charge conservation would 
imply $\Phi_{\rb{q}} = -1$.
Figure \ref{fig:phiq} shows the measured $\Phi_{\rb{q}}$ values in central Pb+Pb 
reactions\footnote{40 and 80 $A$GeV: 7\% most central, 158 $A$GeV: 10\% most central} 
at 40, 80, and 158 $A$GeV beam energy. $\Phi_{\rb{q}}$ depends strongly on the ratio of
accepted charged particles to the total number of charged particles 
$\langle N_{\rb{ch}}(\Delta y) \rangle / \langle N_{\rb{ch}} \rangle_{\rb{tot}}$, which is varied
by changing the accepted rapidity window $\Delta y$. However, it is approximately independent
of the beam energy. The solid line in Fig. \ref{fig:phiq} labelled
$\Phi_{\rb{q}}^{\rb{cc}}$ represents the expectation for a system with total net charge zero,
in which the only correlations are due to global charge conservation.
\begin{equation}
\Phi_{\rb{q}}^{\rb{cc}} = \sqrt{1 - \langle N_{\rb{ch}} \rangle /
                                    \langle N_{\rb{ch}} \rangle_{\rb{tot}}} - 1
\end{equation}
In order to enlarge any deviations from this trivial effect, the difference
$\Delta \Phi_{\rb{q}} = \Phi_{\rb{q}} - \Phi_{\rb{q}}^{\rb{cc}}$ is displayed in
Fig. \ref{fig:dphiq}. It is found that $\Delta \Phi_{\rb{q}}$ is close to zero.

To study the sensitivity of the $\Delta \Phi_{\rb{q}}$ measure a model, 
describing a quark gluon plasma, was investigated \cite{Zara01}. This model assumes an
ideal gas of massless quarks and gluons in equilibrium
with zero baryonic chemical potential ($\mu_{\rb{B}} = 0$).
The requirement of entropy and local charge conservation during hadronization
allows to extract predictions for $\Delta \Phi_{\rb{q}}$ in different scenarios.
Fig. \ref{fig:dphiq} includes two extreme cases: The one labelled ``Frozen QGP''
assumes hadronization only into pions and no diffusion of the net charge in rapidity space, 
so that the initial QGP-like fluctuations
are conserved. In fact, this results in $\Delta \Phi_{\rb{q}}$ values clearly 
below zero. In the other scenario (``Resonances'') hadronization is happening 
entirely into $\rho$-mesons. It turns out that in this case the initial 
fluctuations are completely obscured by the subsequent decay of the resonances,
which cause a smearing of the original QGP fluctuations in rapidity space.

\subsection{$\langle p_{\rb{t}} \rangle$ fluctuations}
\vspace{-0.6cm}

\begin{figure}[h]
\begin{minipage}[b]{54mm}
\caption{$\Phi_{\rb{pt}}$ for forward rapidities (4.0 $< y <$ 5.5) 
as a function of $\langle N_{\rb{part}} \rangle$. 
Included are p+p, C+C, and Si+Si reactions (filled symbols), as well as 
centrality selected Pb+Pb reactions (open symbols), all at 158 $A$GeV.}
\label{fig:phipt}
\end{minipage}
\hspace{\fill}
\begin{minipage}[b]{100mm}
\begin{center}
\includegraphics[height=100mm]{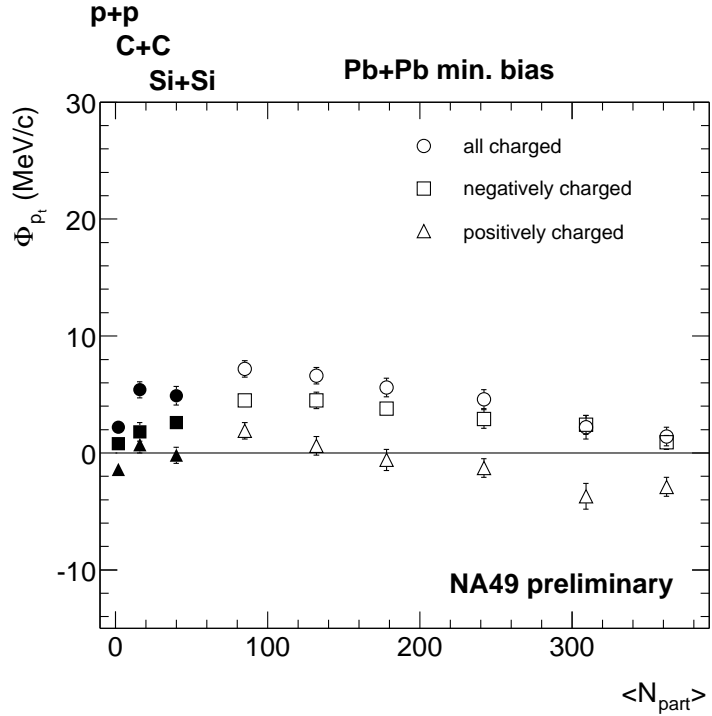}
\end{center}
\vspace{-1.2cm}
\end{minipage}
\end{figure}

For the study of event-by-event $\langle p_{\rb{t}} \rangle$ fluctuations 
again the $\Phi_{\rb{x}}$ observable, defined in Eq.~\ref{eq:phix}, is used.
This time, however, $x$ is replaced by the $p_{\rb{t}}$ of a given particle.
Independent particle emission will again result in $\Phi_{\rb{pt}} = 0$.
The $\Phi_{\rb{pt}}$ values, shown in Fig.~\ref{fig:phipt}, are all corrected
for the effects of the two track resolution of the detector.
Generally, it is found that $\Phi_{\rb{pt}}$ is small for all investigated
reaction systems: $|\Phi_{\rb{pt}}| < 10 \;\rbb{MeV}/c$. A weak centrality 
dependence is observable, with a maximum in $\Phi_{\rb{pt}}$ for very
peripheral Pb+Pb collisions. Also, $\Phi_{\rb{pt}}$ is clearly charge dependent:
$\Phi_{\rb{pt}}$ is always larger for negatively charged particles than for
positively charged. This may be a reflection of the fact that the positively
charged particles contain a larger fraction of baryons, which are subject to
Fermi-Dirac statistics, while the negatively charged particles are dominated
by bosons \cite{Mrow98}. 
However, $\Phi_{\rb{pt}}$ for all charged particles is still higher than
for the negatively charged ones, indicating that there are additional
correlations present.
\vspace{-0.6cm}
\begin{figure}[htb]
\begin{center}
\includegraphics[height=120mm]{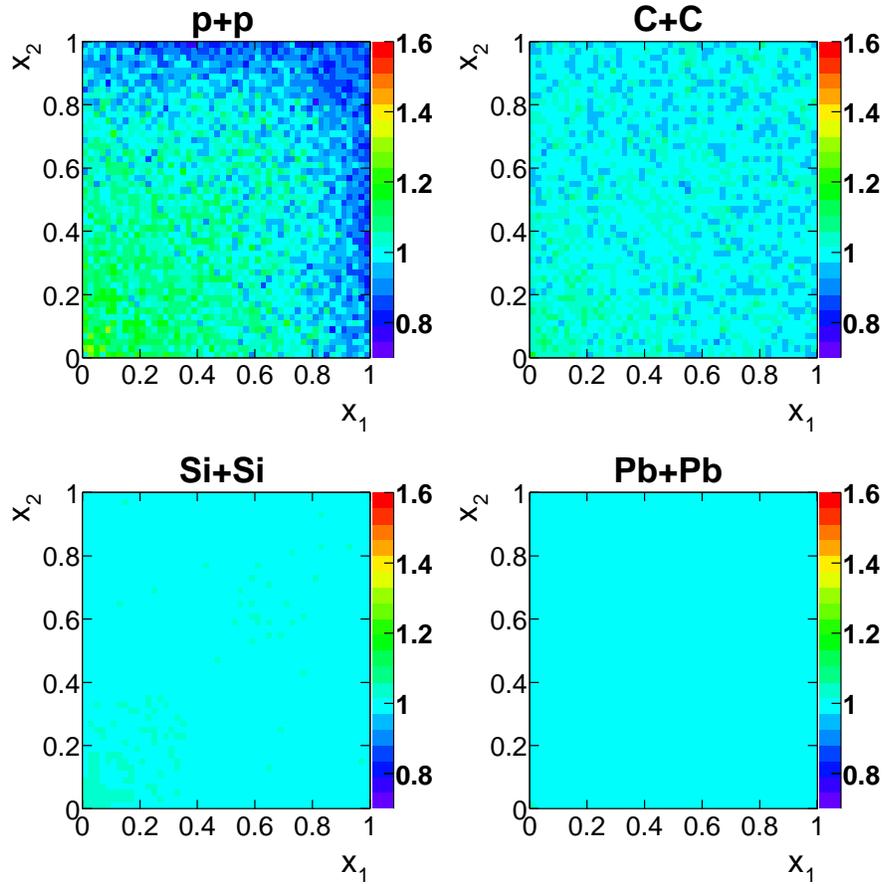}
\end{center}
\vspace{-1.2cm}
\caption{Two particle correlations for p+p, C+C, Si+Si, and central Pb+Pb
reactions at 158 $A$GeV. All figures are on the same scale.}
\label{fig:x1x2}
\end{figure}

Since $\Phi_{\rb{pt}}$ is a global observable, a small value of $\Phi_{\rb{pt}}$ 
does not neccessarily imply the absence of any strong correlation. It is also 
possible that contributions of two effects just cancel out each
other. Looking at two particle correlations provides a more differential way 
of studying $\langle p_{\rb{t}} \rangle$ fluctuations \cite{Trai00}. 
In the procedure employed here, first the $p_{\rb{t}}$ of a given particle 
is transformed into a cumulative variable $x$ \cite{Bial90}:
\begin{eqnarray}
x(p_{\rb{t}}) = \frac{ \int_{0}^{p_{\rb{t}}} \frac{dn}{dp^{\prime}_{\rb{t}}} 
                       dp^{\prime}_{\rb{t}} }
                     { \int_{0}^{\infty}     \frac{dn}{dp^{\prime}_{\rb{t}}} 
                       dp^{\prime}_{\rb{t}} } 
&
\rbb{, where} 
\;dn/dp^{\prime}_{\rb{t}}\; 
\rbb{is the inclusive} 
\;p_{\rb{t}}\;
\rbb{distribution.} 
&
\end{eqnarray}
Then two particle correlation plots are generated by plotting $x_{\rb{1}}$ versus
$x_{\rb{2}}$ for all particle pairs inside one event. Figure~\ref{fig:x1x2} shows
the result for different reactions systems at 158 $A$GeV. 
While there is a clear structure visible
in p+p reactions, reflecting the long range correlations present in this case, these
structures get more and more diluted when going to larger systems. On one side this
is naturally due to the effect of the increased combinatorics between the growing
number of particle pairs, an effect that is removed by the $\Phi_{\rb{pt}}$ measure.
On the other side differences in the reaction dynamics
between elementary p+p and nucleus-nucleus collisions will show up in the two
particle correlations.
The relation between $\Phi_{\rb{pt}}$ and the two particle correlations is still
under study.
\vspace{-0.6cm}
\begin{figure}[htb]
\begin{center}
\includegraphics[height=60mm]{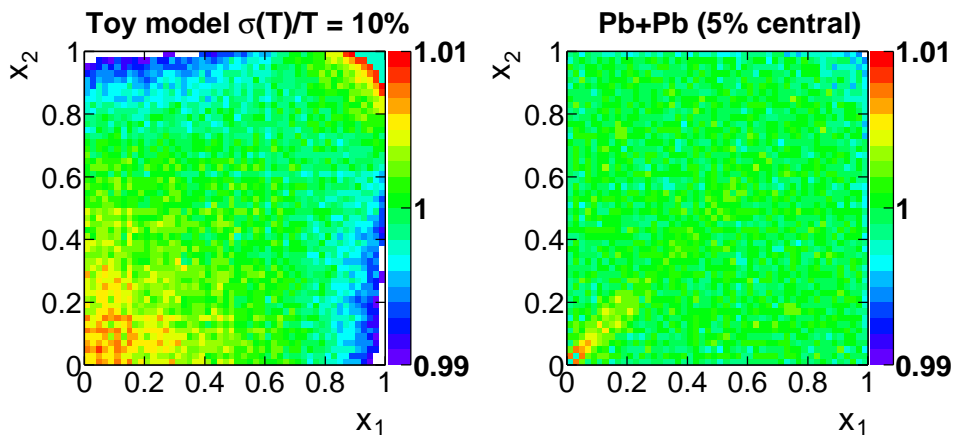}
\end{center}
\vspace{-1.2cm}
\caption{Comparison of a temperature fluctuation model (left hand side) with 
$\sigma(T)/T = 10 \%$ to central Pb+Pb data at 158 $A$GeV (right hand side).
Please note that the data are on a different scale than in Fig.~\ref{fig:x1x2}.}
\label{fig:xsim}
\end{figure}

To see how dynamical fluctuations affect the two particle correlations, a model
study is performed. In this model the only source of fluctuations are the
event-by-event fluctuations of the slope parameter $T$ of the transverse 
momentum spectra. As can be seen on the left panel of Fig.~\ref{fig:xsim},
$T$ fluctuations of the order of 10\% already result in a very prominent 
structure in the two particle correlation. A structure on this level is
clearly absent in the central Pb+Pb data (right panel of Fig.~\ref{fig:xsim}).
Here only short range correlations (e.g. Bose Einstein correlations) are 
visible as an enhancement close to the diagonal.
A comparison of the measured $\Phi_{\rb{pt}}$ in the 5\% most central Pb+Pb
reactions at 158 $A$GeV to a prediction for small $T$ fluctuations \cite{Koru01}
actually suggests that $\sigma(T)/T$ is smaller than 1\%.

\section{Correlations}

\subsection{Centrality dependence of h$^{-}$h$^{-}$ Bose Einstein correlations}

\begin{figure}[htb]
\begin{center}
\includegraphics[height=75mm]{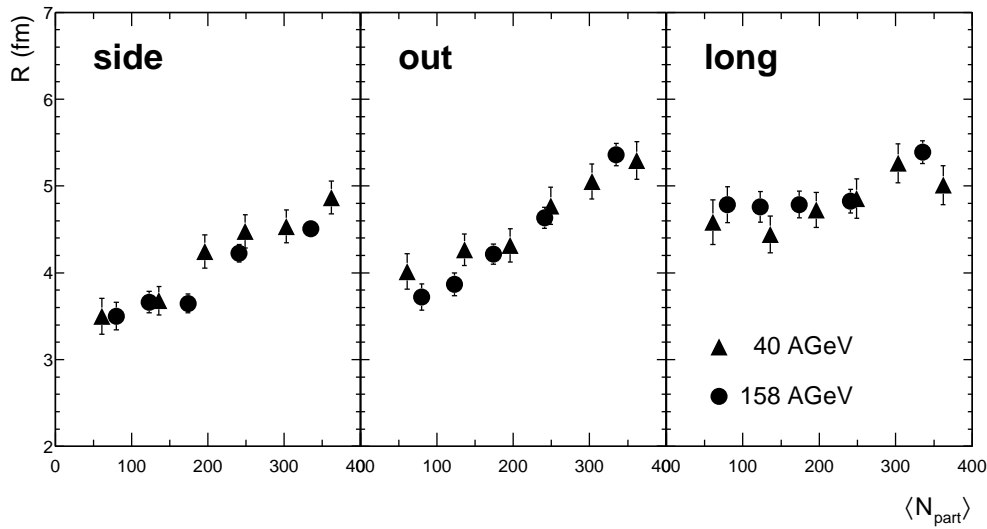}
\end{center}
\vspace{-1.2cm}
\caption{The radius parameters as a function of $\langle N_{\rb{part}} \rangle$ for
centrality selected Pb+Pb reactions at 40 and 158 $A$GeV.}
\label{fig:pionhbt}
\end{figure}

The radius parameters shown in Fig.~\ref{fig:pionhbt} are derived from the fit of the 
Bertsch-Pratt parametrization in the LCMS\footnote{Longitudinally Co-Moving System.}
\begin{equation}
C_{\rb{BP}} = 1 + \lambda \exp( -   R_{\rb{side}}^{2}     \, q_{\rb{side}}^{2}
                                -   R_{\rb{out}}^{2}      \, q_{\rb{out}}^{2}
                                -   R_{\rb{long}}^{2}     \, q_{\rb{long}}^{2}
                                - 2 R_{\rb{out long}}^{2} \, q_{\rb{out}} q_{\rb{long}} )
\end{equation}
to the h$^{-}$h$^{-}$ correlation function (c.f.).
The Coulomb correction is included in the fit
procedure, and is applied only to the fraction of real pairs in the c.f.
For both beam energies the pairs are in the c.m. rapidity region $0.0 < y^{*} < 0.5$ with
$\langle k_{\rb{t}} \rangle = 180 \: \rbb{MeV}/c$.
The radius parameters in side and out direction show a significant increase with the 
system size. $R_{\rb{long}}$, however, shows no clear evidence for a variation with
$\langle N_{\rb{part}} \rangle$, except perhaps for very central reactions. 
The results at 40 and 158 $A$GeV are very similar.

\subsection{Kaon Bose Einstein correlations in central Pb+Pb reactions}

\begin{figure}[htb]
\begin{center}
\includegraphics[height=90mm]{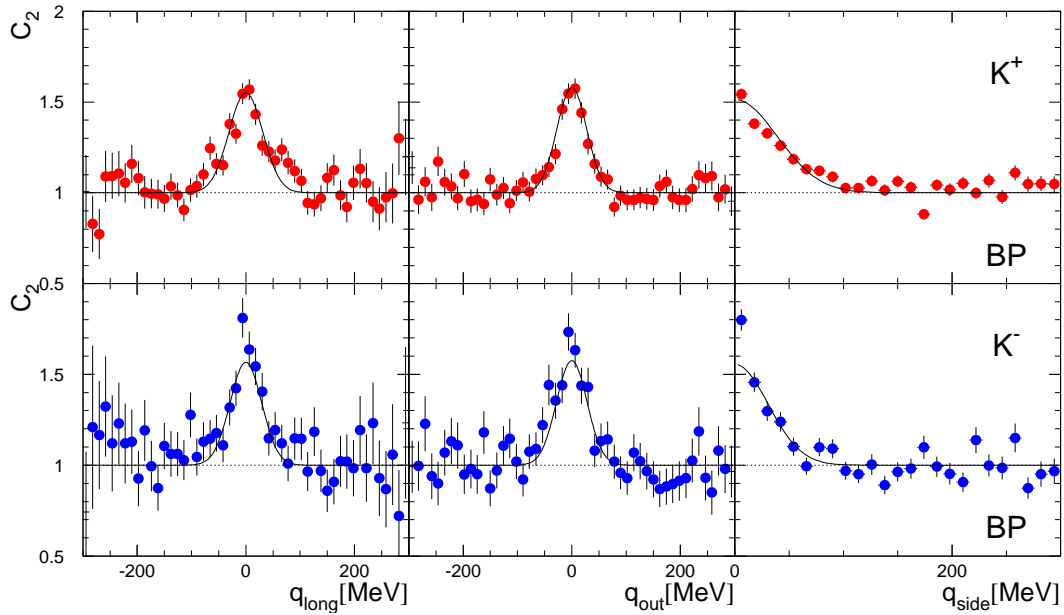}
\end{center}
\vspace{-1.2cm}
\caption{The correlation functions of charged kaons for the 5\% most central
Pb+Pb reactions at 158 $A$GeV close to mid-rapidity ($0.0 < y^{*} < 0.5$).}
\label{fig:kaonhbt}
\end{figure}

\vspace{0.3cm}
\begin{tabular}{|l|c|c|c|}                                                                  \hline
    & K$^{+}$ (NA49)           & K$^{-}$ (NA49)           & K$^{+}$ (NA44) \cite{NA4401} \\ \hline
 $\langle m_{\rb{t}} \rangle$ (GeV) 
    & 0.62                     & 0.61                     & 0.51                         \\ \hline 
 $R_{\rb{side}}$ (fm)               
    & $3.58 \pm 0.40 \pm 0.28$ & $4.55 \pm 0.31 \pm 0.39$ & $4.04 \pm 0.28 \pm 0.32$     \\ \hline
 $R_{\rb{out}}$ (fm)                
    & $5.07 \pm 0.27 \pm 0.35$ & $4.97 \pm 0.39 \pm 0.33$ & $4.12 \pm 0.26 \pm 0.31$     \\ \hline
 $R_{\rb{long}}$ (fm)               
    & $4.46 \pm 0.25 \pm 0.39$ & $4.78 \pm 0.33 \pm 0.40$ & $4.36 \pm 0.33 \pm 0.32$     \\ \hline  
\end{tabular}
\newline

Figure~\ref{fig:kaonhbt} shows the c.f. of charged kaons together 
with the applied fit of the Bertsch Pratt parametrization.
The data points are corrected for the Coulomb interaction 
and the momentum resolution. 

The table summarizes the values of the radius parameters together with statistical and 
systematical errors. Also included are published results from the NA44 collaboration 
\cite{NA4401} on K$^{+}$ correlations that agree quite well with our measurement.

\subsection{Other two particle correlations}

Apart from the effect of quantum statistics, the c.f. reflects
also the influence of the final state interaction. This can be exploited
to gain useful information from correlations between non identical particles.
Results on $\pi$p correlations allow to study relative space time
asymmetries \cite{Ledn96}, and $\Lambda$p correlation give access to the 
source size \cite{Wang99}.
Additionally, one can use the c.f. as a tool to learn about
the two particle interaction in cases where it is unknown, like in the 
$\Lambda \Lambda$ case \cite{Grei89}.
The above correlations are studied in the variable:
\begin{eqnarray}
Q = q_{\rb{inv}} = 2 k^{*} & 
\rbb{with} & 
k^{*} = \frac{1}{2} (\vec{p}_{\rb{1}} - \vec{p}_{\rb{2}}) \:\:
\rbb{(in pair c.m.).}
\end{eqnarray}

\subsubsection{$\pi$p correlations}

\begin{figure}[h]
\begin{minipage}[b]{54mm}
\caption{The ratio $R_{\rb{+}}/R_{\rb{-}}$ in out direction for the 20\% most 
central Pb+Pb reactions at 158 $A$GeV.}
\label{fig:pipasym}
\end{minipage}
\hspace{\fill}
\begin{minipage}[b]{100mm}
\begin{center}
\includegraphics[height=90mm]{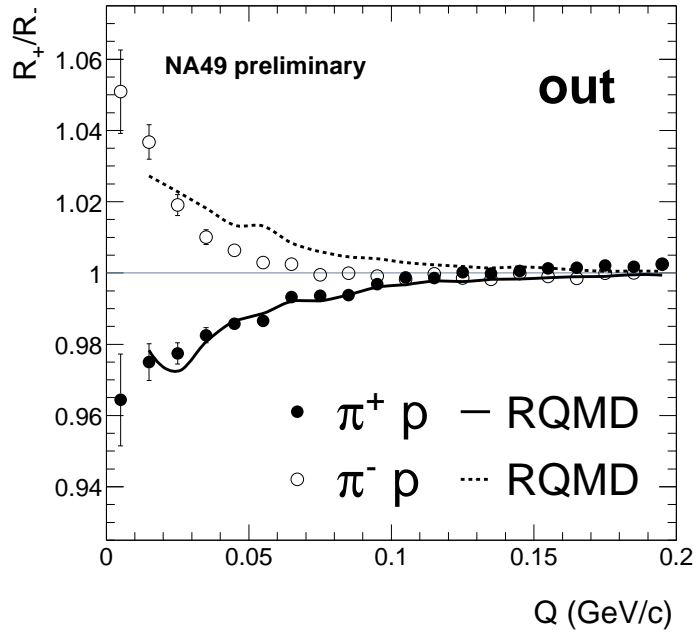}
\end{center}
\vspace{-1.2cm}
\end{minipage}
\end{figure}

Since the relative wave function of pairs of non identical particles contains
odd terms in $\vec{k}^{*} \cdot \vec{r}^{*}$, where $\vec{r}^{*}$ are the
emission points in the pair c.m., relative space time asymmetries
should become visible in the ratio $R_{\rb{+}}/R_{\rb{-}}$. Here, $R_{\rb{+}}$ is the
c.f. containing all pairs with $\vec{k}_{\rb{out}}^{*} \cdot \vec{r}^{*} > 0$, while $R_{\rb{-}}$
consists of pairs with $\vec{k}_{\rb{out}}^{*} \cdot \vec{r}^{*} < 0$, where the 
out-direction is defined in the LCMS.
As can be seen from Fig.~\ref{fig:pipasym} a clear asymmetry is observed in the
data, which goes into opposite directions for $\pi^{+}$p and $\pi^{-}$p pairs.
This mirror symmetry is caused by the fact that the aymmetry in $R_{\rb{+}}/R_{\rb{-}}$
is effected mainly by the Coulomb interaction, which introduces a dependence on
the charge sign.
Additionally, Fig.~\ref{fig:pipasym} includes the result from a RQMD simulation.
Due to its long range character the Coulomb interaction is very sensitive to
the tails in the spatial distribution of the source. 
The simulated emission points have been scaled by a factor of 0.8 in accordance 
with the analysis of $\pi^{+}\pi^{-}$ and $\pi^{+}/\pi^{-}$p correlation functions.

\subsubsection{$\Lambda$p correlations}

\begin{figure}[h]
\begin{minipage}[b]{77mm}
\begin{center}
\includegraphics[height=75mm]{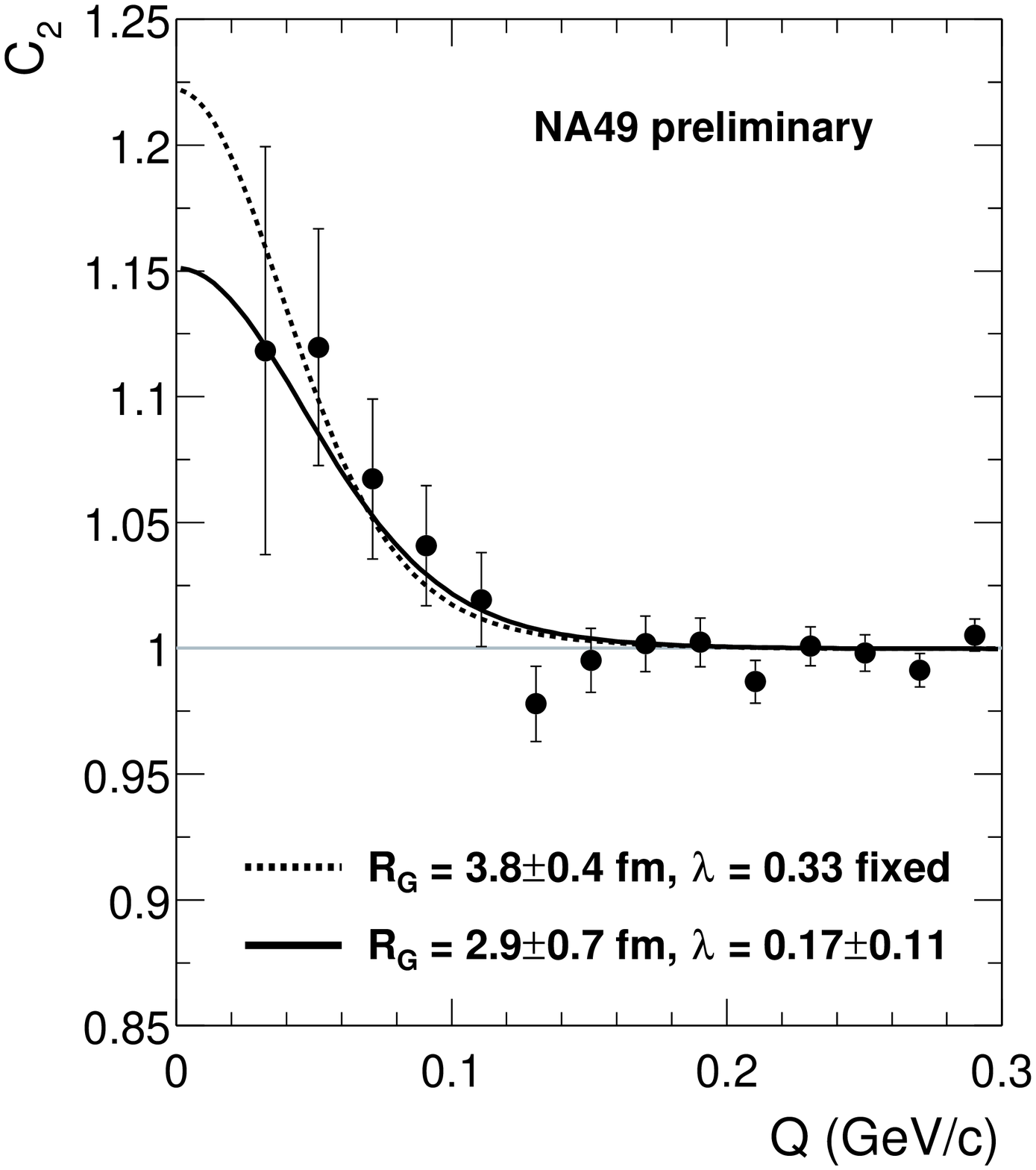}
\end{center}
\vspace{-1.2cm}
\caption{The $\Lambda$p correlation function for the 20\% most central Pb+Pb
reactions at 158 $A$GeV. The lines represent fits of the calculated c.f. with
fixed $\lambda$ parameter (dashed) and and free $\lambda$ (solid).}
\label{fig:plam}
\end{minipage}
\hspace{\fill}
\begin{minipage}[b]{77mm}
\begin{center}
\includegraphics[height=75mm]{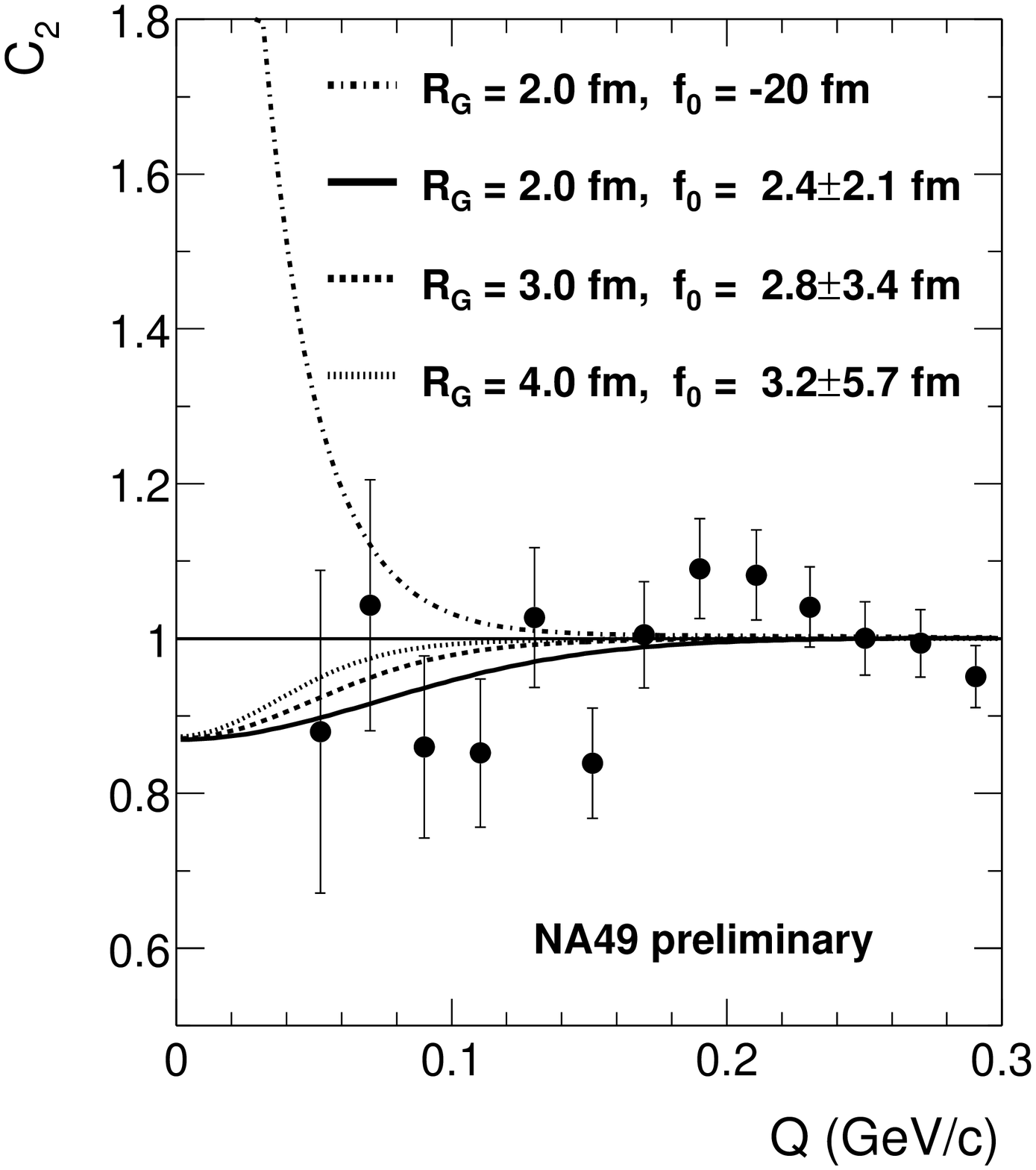}
\end{center}
\vspace{-1.2cm}
\caption{The $\Lambda\Lambda$ correlation function for the 20\% most central
Pb+Pb reactions. The lines display the fit results of the calculated c.f. to the data
for different fixed Gaussian source radii $R_{G}$.}
\label{fig:lamlam}
\end{minipage}
\end{figure}

Figure~\ref{fig:plam} shows the measured c.f., containing 60k pairs
with $Q < 0.3 \:\rbb{GeV}/c$, together with theoretical c.f. fitted to the data.
The calculation is based on an effective range approximation, using a s-wave
scattering length of $f_{\rb{0}} =$ -2.3 fm (singlet) / -1.8 fm (triplet) \cite{Ledn82}.
The Gaussian source is assumed to be spherically symmetric and static, and is defined by
the parameter $R_{\rb{G}}$. Figure~\ref{fig:plam} includes two different fits. In the first
$R_{\rb{G}}$ and the $\lambda$ parameter can vary freely. In the second
$\lambda$ is fixed to a value that is estimated from the background of particle
misidentification and the contribution from feed down. Both fits suggest a Gaussian
source size of $R_{\rb{G}}$ = 3 - 4 fm, which is compatible with the NA49 result on
pp correlations ($R_{\rb{G}} = 4.0 \pm 0.15 ^{+ 0.06}_{-0.18} \: \rbb{fm}$) \cite{NA4999}. 

\subsubsection{$\Lambda\Lambda$ correlations}

The significance of the measured $\Lambda\Lambda$ c.f., shown in Fig.~\ref{fig:lamlam}, 
is unfortunately limited by low statistics (3500 pairs with $Q < 0.3 \:\rbb{GeV}/c$)
and does not show any clear structure.
Nevertheless, it is worthwhile to do a comparison to theoretical expectations, in an
attempt to limit the range of possible parameter values. Therefore, a fit is performed
where $R_{\rb{G}}$ and $\lambda$ are now fixed and the scattering length $f_{\rb{0}}$,
describing the strengh of the interaction, is varied. The result indicates that the
c.f. would favour a relatively small $f_{\rb{0}}$, quite independent from the assumed 
source size. As a comparison a calculation with $f_{\rb{0}}$ = -20 fm is also shown,
which would correspond to the scattering length in the nucleon-nucleon case, and
which looks rathers unlikely, although is is not totally ruled out.

\subsection{Deuteron coalescence}

\begin{figure}[h]
\begin{minipage}[b]{77mm}
\begin{center}
\includegraphics[height=75mm]{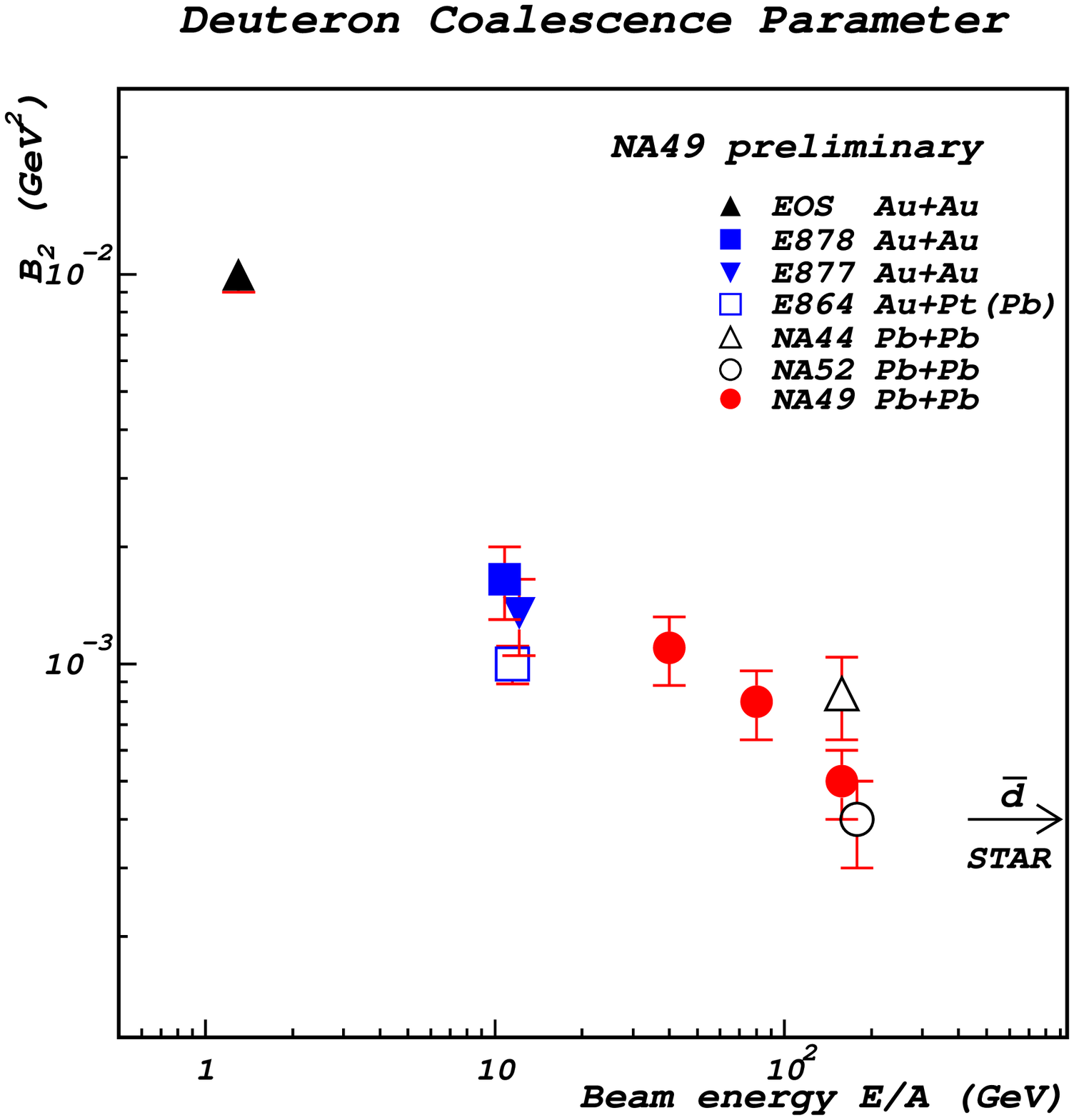}
\end{center}
\vspace{-1.2cm}
\caption{The deuteron coalescence factor $B_{\rb{2}}$ as a function of the beam energy
for central nucleus-nucleus reactions.}
\label{fig:b2energy}
\end{minipage}
\hspace{\fill}
\begin{minipage}[b]{77mm}
\begin{center}
\includegraphics[height=75mm]{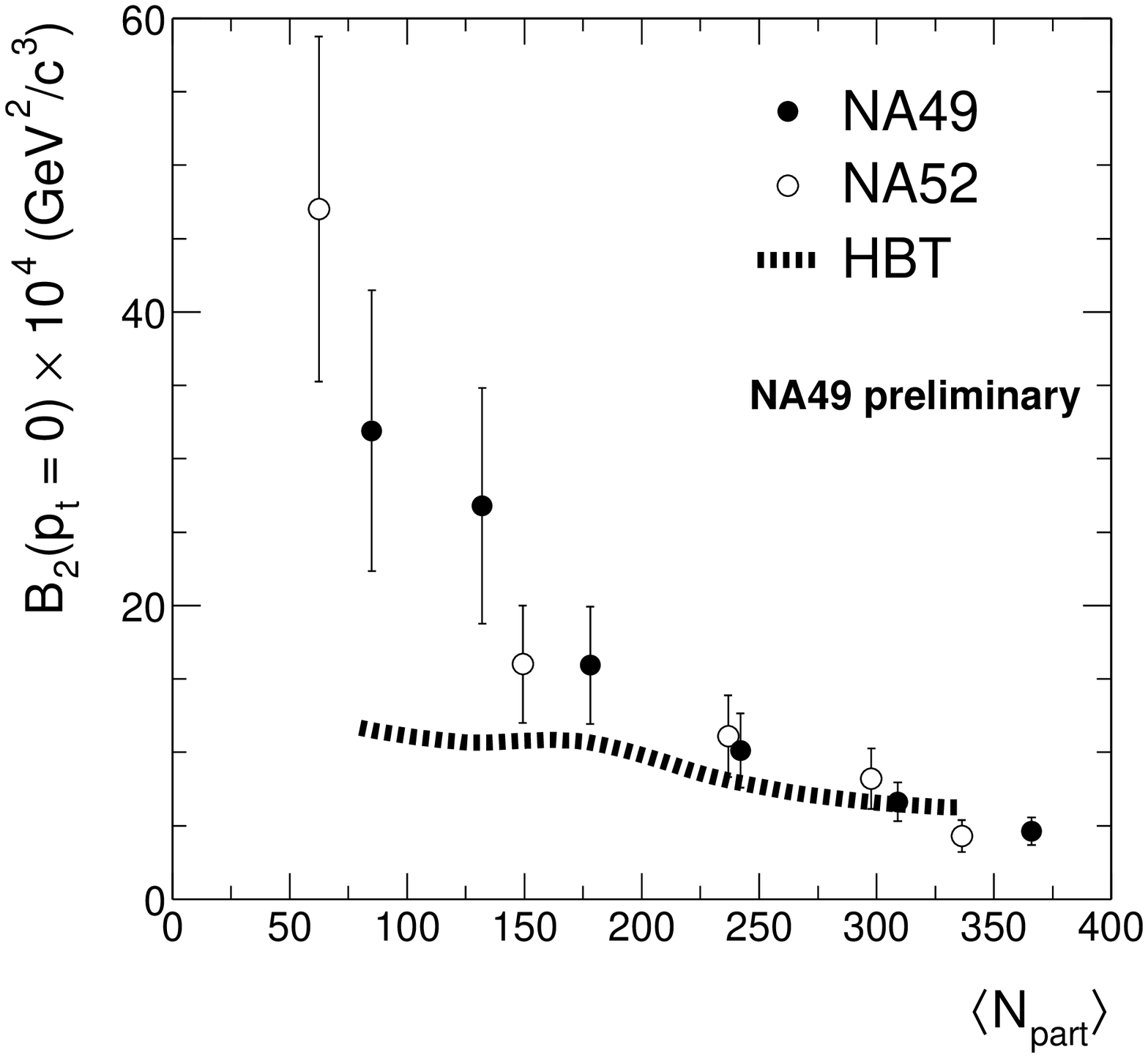}
\end{center}
\vspace{-1.2cm}
\caption{The deuteron coalescence factor $B_{\rb{2}}$ as a function of the system size
for minimum bias Pb+Pb reactions at 158 $A$GeV.}
\label{fig:b2size}
\end{minipage}
\end{figure}

From the measurement of proton and deuteron spectra (all at mid-rapidity for central
reactions\footnote{40 and 80 $A$GeV: 7\% most central, 158 $A$GeV: 5\% most central}),
a deuteron coalescence factor $B_{\rb{2}}$ can be derived:
\begin{equation}
  E_{\rb{d}} \frac{d^{3} N_{\rb{d}}}{dp^{3}_{\rb{d}}} =  
  B_{\rb{2}}
  \left( E_{\rb{p}} \frac{d^{3} N_{\rb{p}}}{dp^{3}_{\rb{p}}}  \right)^{2}
  \: , \:
  p_{\rb{d}} = 2 p_{\rb{p}} 
\end{equation}
Figure~\ref{fig:b2energy} displays the extracted $B_{\rb{2}}$ for different beam energies,
also including data from the AGS and other CERN SPS experiments. $B_{\rb{2}}$ decreases significantly
with increasing beam energy. Even in the SPS energy range, taken alone, 
there is a change by a factor of 2.
$B_{\rb{2}}$ is also strongly dependent on the system size, as is illustrated in Fig.~\ref{fig:b2size}.
Going from very peripheral to central reactions, it decreases by a factor of almost 10.
This is in contrast to the centrality dependence of the radii from Bose Einstein correlations 
(see Fig.~\ref{fig:pionhbt}), which can be related to $B_{\rb{2}}$ via \cite{Llop95}:
\begin{eqnarray}
  B_{\rb{2}} = \frac{3}{4} (\sqrt{\pi} \hbar c)^{3} \frac{m_{\rb{d}}}{m_{\rb{p}}^{2}} 
                                                    \frac{1}{R_{\rb{G}}^{3}}      &
  \rbb{with} &
  R_{\rb{G}} = \sqrt[3]{R_{\rb{side}}^{2} R_{\rb{long}}}
\end{eqnarray}
As is demonstrated in Fig.~\ref{fig:b2size} the $B_{\rb{2}}$ derived from the HBT volume 
changes much less with the centrality than the measured deuteron coalescence parameter.

\vspace{0.6cm}
\begin{footnotesize}
Acknowledgements: This work was supported by the Director, Office of Energy Research,
Division of Nuclear Physics of the Office of High Energy and Nuclear Physics
of the US Department of Energy (DE-ACO3-76SFOOO98 and DE-FG02-91ER40609),
the US National Science Foundation,
the Bundesministerium fur Bildung und Forschung, Germany,
the Alexander von Humboldt Foundation,
the UK Engineering and Physical Sciences Research Council,
the Polish State Committee for Scientific Research (2 P03B 130 23 and 2 P03B 02418),
the Hungarian Scientific Research Foundation (T14920 and T32293),
Hungarian National Science Foundation, OTKA, (F034707),
the EC Marie Curie Foundation,
and the Polish-German Foundation.
\end{footnotesize}


\vspace{0.6cm}

\begin{thebibliography}{9}

\bibitem{NA49NM} S.~Afanasiev et al., Nucl. Instrum. Meth. {\bf A430} (1999), 210.
 
\bibitem{Asak00} M.~Asakawa, U.~Heinz, and B.~M\"uller, Phys. Rev. Lett. {\bf 85} (2000), 2072.

\bibitem{Jeon00} S.~Jeon and V.~Koch, Phys. Rev. Lett. {\bf 85} (2000), 2076. 

\bibitem{Gazd92} M.~Ga\'zdzicki and S.~Mr\'owczy\'nski, Z. Phys. {\bf C54} (1992), 127. \\
                 M.~Ga\'zdzicki, Eur. Phys. J. {\bf C8} (1999), 131.

\bibitem{Zara01} J.~Zaranek, hep-ph/0111228, to appear in Phys. Rev. C.

\bibitem{Mrow98} S.~Mr\'owczy\'nski, Phys. Lett. {\bf B439} (1998), 6.

\bibitem{Trai00} T.~Trainor, hep-ph/0001148.

\bibitem{Bial90} A.~Bia{\l}as, M.~Ga\'zdzicki, Phys. Lett. {\bf B252} (1990), 483.

\bibitem{Koru01} R.~Korus, S.~Mr\'owczy\'nski, M.~Rybczy\'nski, and Z.~W{\l}odarczyk,
                 Phys. Rev. {\bf C64} (2001), 054908.

\bibitem{NA4401} I.~G.~Bearden et al., Phys. Rev. Lett. {\bf 87} (2001), 112301.

\bibitem{Ledn96} R.~Lednick\'y, V.~L.~Lyuboshitz, B.~Erazmus, and D.~Nouais,
                 Phys. Lett. {\bf B373} (1996), 30.

\bibitem{Wang99} F.~Wang and S.~Pratt, Phys. Rev. Lett. {\bf 83} (1999), 3138.

\bibitem{Grei89} C.~Greiner and B.~M\"uller, Phys. Lett. {\bf B219} (1989), 199.

\bibitem{Ledn82} R.~Lednick\'y and V.~L.~Lyuboshitz, Sov. J. Nucl. Phys. {\bf 35} (1982), 770.

\bibitem{NA4999} H.~Appelsh\"auser et al., Phys. Lett. {\bf B467} (1999), 21.

\bibitem{Llop95} W.~Llope et al., Phys. Rev. {\bf C52} (1995), 2004.

\end{thebibliography}
\end{document}